\documentclass[letterpaper,english,aps,prl,superscriptaddress,twocolumn,showpacs,showkeys]{revtex4}
\usepackage[T1]{fontenc}
\usepackage[latin1]{inputenc}
\usepackage{amsmath}
\usepackage{graphicx}

\makeatletter
\usepackage{babel}
\makeatother
\begin{document}

\title{Anti-Stokes laser cooling in bulk Erbium-doped materials}

\author{Joaquín Fernandez}

\email{wupferoj@bi.ehu.es}

\affiliation{Departamento de Física Aplicada I, E.T.S. Ingeniería de Bilbao, Alda.
Urquijo s/n, 48013 Bilbao, Spain}

\affiliation{Unidad Física de Materiales CSIC-UPV/EHU and Donostia International
Physics Center, Apartado 1072, 20080 San Sebastian, Spain}

\author{Angel J. Garcia--Adeva}

\affiliation{Departamento de Física Aplicada I, E.T.S. Ingeniería de Bilbao, Alda.
Urquijo s/n, 48013 Bilbao, Spain}

\author{Rolindes Balda}

\affiliation{Departamento de Física Aplicada I, E.T.S. Ingeniería de Bilbao, Alda.
Urquijo s/n, 48013 Bilbao, Spain}

\affiliation{Unidad Física de Materiales CSIC-UPV/EHU and Donostia International
Physics Center, Apartado 1072, 20080 San Sebastian, Spain}

\begin{abstract}
We report the first observation of anti-Stokes laser-induced cooling
in the Er$^{3+}$:KPb$_{2}$Cl$_{5}$ crystal and in the Er$^{3+}$:CNBZn
(CdF$_{2}$-CdCl$_{2}$-NaF-BaF$_{2}$-BaCl$_{2}$-ZnF$_{2}$) glass.
The internal cooling efficiencies have been calculated by using photothermal
deflection spectroscopy. Thermal scans acquired with an infrared thermal
camera proved the bulk cooling capability of the studied samples.
Implications of these results are discussed.
\end{abstract}



\maketitle
The basic principle that anti-Stokes fluorescence might be used to
cool a material was first postulated by Pringsheim in 1929. Twenty
years later Kastler suggested \cite{kastler1950} that rare-earth-doped
crystals might provide a way to obtain solid-state cooling by anti-Stokes
emission (CASE). A few years later, the invention of the laser promoted
the first experimental attempt by Kushida and Geusic to demonstrate
radiation cooling in a Nd$^{3+}$:YAG crystal \cite{kushida1968}.
However, it was not until 1995 that the first solid-state CASE was
convincingly proven by Epstein and coworkers in an ytterbium-doped
heavy-metal fluoride glass \cite{Epstein1995}. Since then on, the
efforts to develop other different materials doped with rare-earth
(RE) ions were unsuccessful due to the inherent characteristics of
the absorption and emission processes in RE ions. In most of the materials
studied, the presence of nonradiative (NR) processes hindered the
CASE performance. As a rule of thumb a negligible impurity parasitic
absorption and near-unity quantum efficiency of the anti-Stokes emission
from the RE levels involved in the cooling process are required, so
that NR transition probabilities by multiphonon emission or whatever
other heat generating process remain as low as possible. These constraints
could explain why most of the efforts to obtain CASE in condensed
matter were performed on trivalent ytterbium doped solids (glasses
\cite{Hoyt2003} and crystals \cite{Bowman2000,Medioroz2002}) having
only one excited state manifold which is placed $\sim10000$ cm$^{-1}$
above the ground state. The only exception was the observation of
CASE in a thulium-doped glass by using the transitions between the
$^{3}$H$_{6}$ and $^{3}$H$_{4}$ manifolds to cool down the sample
\cite{Hoyt2000}. Therefore, it is easy to see that identifying new
optically active ions and materials capable of producing CASE is still
an open problem with very important implications from both the fundamental
and practical points of view.

On the other hand, the recent finding of new low phonon materials
(both glasses \cite{Fernandez2000} and crystals \cite{Medioroz2002})
as RE hosts which may significantly decrease the NR emissions from
excited state levels have renewed the interest in investigating new
RE anti-Stokes emission channels. In this work, we present the first
experimental demonstration of anti-Stokes laser-induced cooling in
two different erbium-doped matrices: a low phonon KPb$_{2}$Cl$_{5}$
crystal and a fluorochloride glass. In order to assess the presence
of internal cooling in these systems we employed the photothermal
deflection technique, whereas the bulk cooling was detected by means
of a calibrated thermal sensitive camera. The cooling was obtained
by exciting the Er$^{3+}$ ions at the low energy side of the $^{4}$I$_{9/2}$
manifold with a tunable Ti:sapphire laser. It is worthwhile to mention
that this excited state, where cooling can be induced, is also involved
in infrared to visible upconversion processes nearby the cooling spectral
region \cite{Balda2004}. Moreover, it is also noticeable that the
laser induced cooling can be easily reached at wavelengths and powers
at which conventional laser diodes operate, which renders these systems
very convenient for applications, such as compact solid-state optical
cryo-coolers.

Single crystals of nonhygroscopic Er$^{3+}$:KPb$_{2}$Cl$_{5}$ were
grown in our laboratory by the Bridgman technique \cite{Voda2004}.
The rare earth content was $0.5$ mol\% of ErCl$_{3}$. The fluorochloride
CNBZn glass doped with $0.5$ mol\% of ErF$_{3}$ was synthesized
at the Laboratoire de Verres et Ceramiques of the University of Rennes.
The experimental setup and procedure for photothermal deflection measurements
have been described elsewhere \cite{Fernandez2000,fernandez2001}.
The beam of a tunable cw Ti:sapphire ring laser (Coherent 899), with
a maximum output power of $2.5$ W, was modulated at low frequency
($1-10$ Hz) by a mechanical chopper and focused into the middle of
the sample with a diameter of $\sim100$ $\mu$m. The copropagating
helium-neon probe laser beam ($\lambda=632.8$ nm) was focused to
$\sim60$ $\mu$m, co-aligned with the pump beam, and its deflection
detected by a quadrant position detector. The samples (of sizes $4.5\times6.5\times2.7\,\text{mm}^{3}$
and $10.7\times10.7\times2.2\,\text{mm}^{3}$ for the crystal and
glass, respectively) were freely placed on a teflon holder inside
a low vacuum ($\sim10^{-2}$ mbar) cryostat chamber at room temperature.

The cooling efficiencies of the Er$^{3+}$-doped materials were evaluated
at room temperature by measuring the quantum efficiency (QE) of the
emission from the $^{4}$I$_{9/2}$ manifold in the heating and cooling
regions by means of the photothermal deflection spectroscopy in a
collinear configuration \cite{Fernandez2000,fernandez2001}. The evaluation
of the QE has been carried out by considering a simplified two level
system for each of the transitions involved. In the photothermal collinear
configuration, the amplitude of the angular deviation of the probe
beam is always proportional to the amount of heat the sample exchanges,
whatever its optical or thermal properties are. The QE of the transition,
$\eta$, can be obtained from the ratio of the photothermal deflection
amplitude (PDS) to the sample absorption (Abs) obtained as a function
of the excitation wavelength $\lambda$ around the mean fluorescence
wavelength $\lambda_{0}$\begin{equation}
\frac{\text{PDS}}{\text{Abs}}=C\left(1-\eta\frac{\lambda}{\lambda_{0}}\right),\end{equation}
where $C$ is a proportionality constant that depends on the experimental
conditions. The mean fluorescence wavelength, above which cooling
is expected to occur, was calculated by taking into account the branching
ratios for the emissions from level $^{4}$I$_{9/2}$. As expected,
the calculated value is close to that found experimentally for the
transition wavelength at which the cooling region begins.%
\begin{figure}
\includegraphics{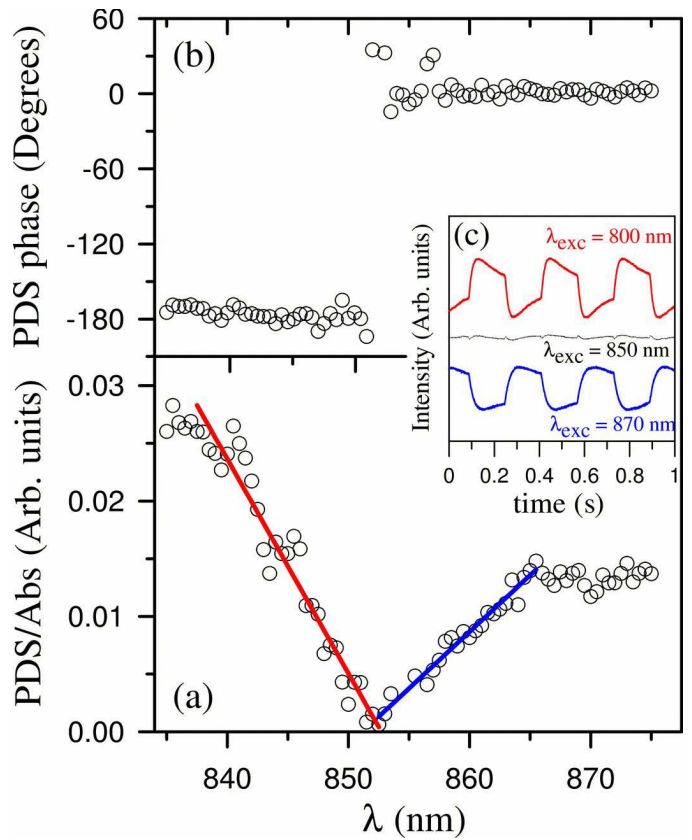}

\caption{\label{fig_pds_crystal}(a) Signal deflection amplitude normalized
by the sample absorption as a function of pumping wavelength for the
Er$^{3+}$:KPb$_{2}$Cl$_{5}$ crystal. (b) Phase of the photothermal
deflection signal as a function of pumping wavelength. (c) Photothermal
deflection signal waveforms in the heating (800 nm) and cooling (870
nm) regions and around the cooling threshold (850 nm).}
\end{figure}

\begin{figure}
\includegraphics{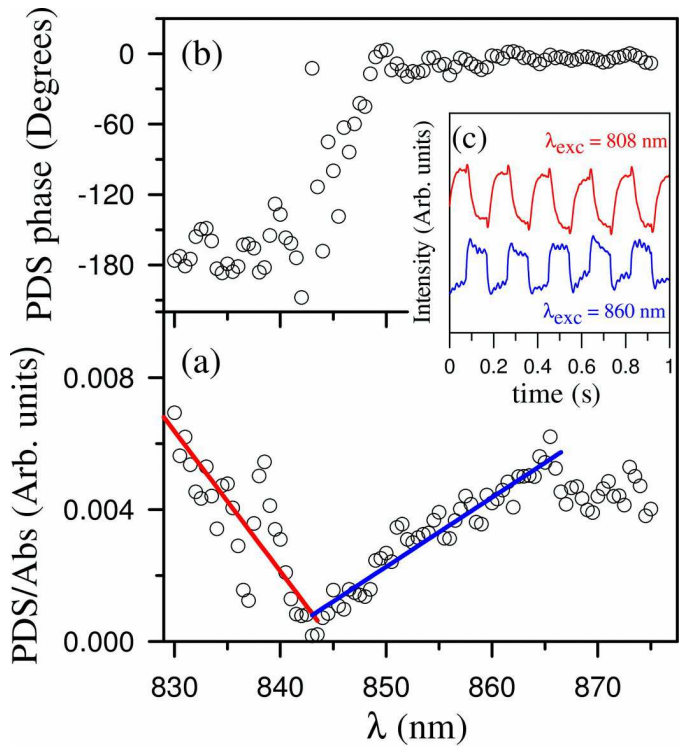}

\caption{\label{fig_pds_glass}(a) Signal deflection amplitude normalized
by the sample absorption as a function of pumping wavelength for the
Er$^{3+}$:CNBZn glass. (b) Phase of the photothermal deflection signal
as a function of pumping wavelength. (c) Photothermal deflection signal
waveforms in the heating (808 nm) and cooling (860 nm) regions.}
\end{figure}

Figure \ref{fig_pds_crystal}a shows the normalized PDS spectrum of
Er$^{3+}$:KPb$_{2}$Cl$_{5}$ crystal around the zero deflection
signal ($852.5$ nm) --obtained at an input power of $1.5$ W-- together
with the best least-squares fitting in both the heating and cooling
regions. The resulting QE values are $0.99973$ and $1.00345$, respectively
and, therefore, the cooling efficiency estimated by using the QE measurements
is $0.37\%$. As predicted by the theory \cite{Jackson1981}, a sharp
jump of $180^{\circ}$ in the PDS phase measured by lock-in detection
can be observed during the transition from the heating to the cooling
region (see Fig.~\ref{fig_pds_crystal}b). The Figure \ref{fig_pds_crystal}c
shows the PDS amplitude waveforms registered in the oscilloscope at
three different excitation wavelengths: $800$ nm (heating region),
$852.5$ (mean fluorescence wavelength), and $870$ nm (cooling region).
As can be noticed, at $852.5$ nm the signal is almost zero whereas
in the cooling region, at $870$ nm, the waveform of the PDS signal
shows an unmistakable phase reversal of $180^{\circ}$ when compared
with the one at $800$ nm. Figure \ref{fig_pds_glass} shows the CASE
results for the Er$^{3+}$:CNBZn glass (obtained at a pump power of
$1.9$ W) where the zero deflection signal occurs around $843$ nm.
The $180^{\circ}$ change of the PDS phase is also clearly attained
but with a little less sharpness than for the Er$^{3+}$:KPb$_{2}$Cl$_{5}$
crystal (see Fig.~\ref{fig_pds_glass}b). The QE values corresponding
to the heating and cooling regions are $0.99764$ and $1.00446$,
respectively, and the estimated cooling efficiency is $0.68\%$. The
PDS waveforms corresponding to the heating and cooling regions are
shown in Fig.~\ref{fig_pds_glass}c. It is worthy to notice that
the cooling processes in both systems can be obtained at quite low
power excitations. As an example, for the Er$^{3+}$:KPb$_{2}$Cl$_{5}$
crystal, CASE is still efficient at a pump power of only $500$ mW.

The results described in the previous paragraphs clearly demonstrate
that these systems are capable of internal laser cooling in a certain
spectral range and even at small pumping powers. We also conducted
measurements of the absolute temperature of the present materials
as a function of time for several pumping powers between $0.25$ and
$1.9$ W and wavelengths in both the heating and cooling regions described
above in order to assess quantitatively their cooling potential. To
perform these measurements, a Thermacam SC 2000 (FLIR Systems) infrared
thermal camera was used. This camera operates between $-40^{\circ}$C
and $500^{\circ}$C object temperature with a precision of $\pm0.1^{\circ}$C.
The detector is an array of $320\times240$ microbolometers. The camera
is connected to an acquisition card interface that is able to record
thermal scans at a rate of $50$ Hz. The absolute temperature was
calibrated with a thermocouple located at the sample holder.

\begin{figure}
\includegraphics{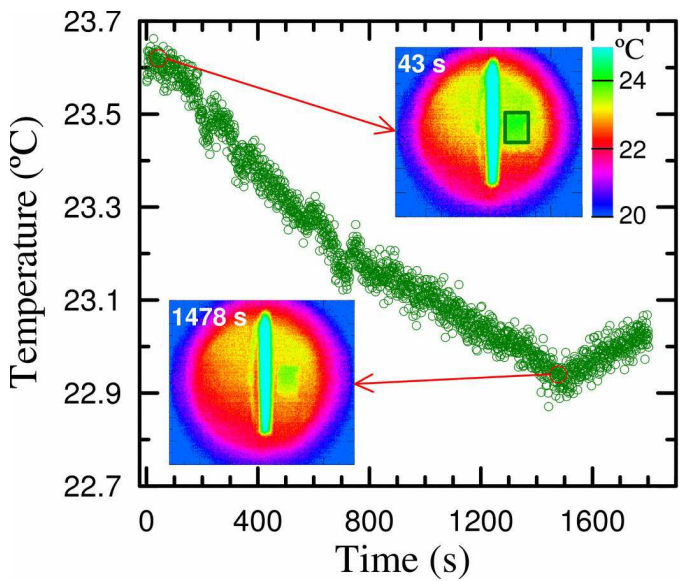}

\caption{\label{fig_temp_crystal}Time evolution of the average temperature
of the Er$^{3+}$:KPb$_{2}$Cl$_{5}$ at 870 nm. The insets show colormaps
of the temperature field of the whole system (sample plus cryostat)
at two different times as measured with the thermal camera. The rectangle
in the upper inset delimits the area used for calculating the average
temperature of the sample.}
\end{figure}
\begin{figure}
\includegraphics{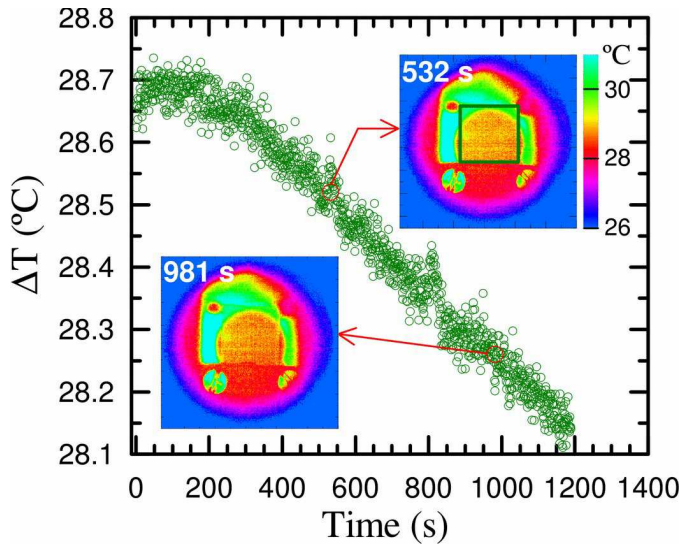}

\caption{\label{fig_temp_glass}Time evolution of the average temperature
of the Er$^{3+}$:CNBZn at 860 nm. The insets show colormaps of the
temperature field of the whole system (sample plus cryostat) at two
different times as measured with the thermal camera. The rectangle
in the upper inset delimits the area used for calculating the average
temperature of the sample.}
\end{figure}

Thermal scans at a rate of $1$ image per second were acquired for
time intervals that depend on the particular data series. The camera
was placed $12$ cm apart from the window cryostat so that a lens
with a field of vision of $45^{\circ}$ allows focusing the camera
on the sample. Figures \ref{fig_temp_crystal} and \ref{fig_temp_glass}
show the runs performed at 870 and 860 nm for the Er$^{3+}$:KPb$_{2}$Cl$_{5}$
crystal and Er$^{3+}$:CNBZn glass samples, respectively, using the
same pump geometry conditions as the ones described above. The laser
power on the sample was fixed at 1.9 W in both cases. According to
the PDS measurements reported above, these pumping wavelengths are
well inside the cooling region for both materials. The insets in Figs.~\ref{fig_temp_crystal}
and \ref{fig_temp_glass} depict some examples of the thermal scans
obtained with the infrared camera. It is clear from those colormaps
that the sample is cooling down as time goes by. However, it is difficult
to extract any quantitative information about the amount the sample
is cooling, as these changes are small compared with the absolute
temperature of its surroundings. For this reason, in order to assess
whether cooling is occurring in the bulk, we calculated the average
temperature of the area enclosed in the green rectangles depicted
in the upper insets of figures \ref{fig_temp_crystal} and \ref{fig_temp_glass}
and the corresponding results constitute the green curves in those
figures. As it is easy to see, both samples cool down under laser
irradiation. The Er$^{3+}$:KPb$_{2}$Cl$_{5}$ sample temperature
drops by $0.7\pm0.1^{\circ}$C in $1500$ s ($30$ minutes). To check
that this temperature change was indeed due to laser cooling, the
laser was turned off at that point. This can be easily identified
as an upturn in the curve that represents the evolution of the sample
temperature, which means that this quantity starts to rise as soon
as the laser irradiation is stopped. On the other hand, the temperature
of the Er$^{3+}$: CNBZn glass sample starts to rise when laser irradiation
starts. After $\sim150$ s ($2$ minutes and a half) this tendency
is inverted and the sample starts to cool down. From that point on
(and in approximately $1000$ s), the average temperature of the sample
drops by $0.5\pm0.1^{\circ}$C. Estimations of the expected bulk temperature
change based on microscopic models proposed by Petrushkin, Samartev
and Adrianov \cite{Petrushkin2001} yield values of $-5^{\circ}$C
and $-10^{\circ}$C for the crystalline and glass samples, respectively,
under the experimental conditions described above. The discrepancy
between the theoretical estimates and the present experimental results
can be attributed to partial re-absorption of the anti-Stokes fluorescence,
not taken into account in these models, or additional absorption processes
of the pumping radiation involving excited states, which are known
to be significant in these materials \cite{Balda2004}. In any case,
if one takes into account the minute concentrations of the optically
active ions in the materials studied in this work and the geometry
of the cooling experiment (single pass configuration), we think that
the results described in this paragraph come to show that CASE in
these materials is extremely efficient.

In conclusion, we have demonstrated cooling by anti-Stokes emission
in two materials doped with Er optically active ions by using a combination
of photothermal deflection measurements and time evolution of the
average temperature of the sample acquired with an infrared camera.
In particular, the photothermal deflection measurements clearly show
internal cooling in the two samples analyzed. The cooling efficiencies
are found to be about $0.4\%$ and $0.7\%$ for the crystal and glass
samples, respectively. These figures are remarkable if one takes into
account the fact that the concentration of the optically active ions
in our materials are about $0.5\%$ of Er$^{3+}$ and that our experiments
are performed in a single pass configuration. From a fundamental perspective,
these results are quite important, as this ion comes to engross the
small list of rare earth ions that are amenable to cooling (Yb$^{3+}$
and Tm$^{3+}$ ions being the other two known so far). On the other
hand, the measurements performed by using the infrared camera demonstrate
that the Er$^{3+}$ ions present in the materials are able to refrigerate
these by $0.7$ and $0.5^{\circ}$C for the crystalline and glass
samples, respectively. This result is extremely important from the
applied point of view, as it paves the way to use this ion as an efficient
anti-Stokes emitter for compact solid state optical refrigerators.
Moreover, it opens a wide field of applications related with the possibility
to use CASE to offset the heat generated by the laser operation in
Er$^{3+}$-based fiber lasers --the so called radiation-balanced lasers
\cite{Bowman1999}-- that would allow to use dual wavelength pumping
to take advantage of the cooling processes occurring at a given wavelength.
This technique could allow to scale up the power of Er$^{3+}$ based
fiber lasers. On the other hand, the use of Er$^{3+}$-doped nanoparticles
for bioimaging or phototherapy could also take advantage of a dual
wavelength pumping (at a nearby wavelength) in order to balance the
thermal damage produced in a soft tissue by the infrared pumping wavelength
at which the upconversion process occurs.

\begin{acknowledgments}
This work was supported by the University of the Basque Country (Grant
No. UPV13525/2001). A.J.G.-A. wants to acknowledge financial support
from the Spanish MEC under the {}``Ramón y Cajal'' program.
\end{acknowledgments}

\end{document}